\documentclass[rreprint,amsmath,amssymb,aps,prb,twocolumn,showpacs,superscriptaddress]{revtex4-1}

\usepackage{graphicx}
\begin{document}

% ***** TITLE
\title{Electronic Structure of the Metallic Antiferromagnet PdCrO$_2$ Measured by Angle-Resolved Photoemission Spectroscopy}

% ***** AUTHORS
\author{J.~A. Sobota}
\affiliation{Stanford Institute for Materials and Energy Sciences, SLAC National Accelerator Laboratory, 2575 Sand Hill Road, Menlo Park, CA 94025, USA}
\affiliation{Geballe Laboratory for Advanced Materials, Departments of Physics and Applied Physics, Stanford University, Stanford, CA 94305, USA}

\author     { K. Kim }	
\affiliation{ Department of Physics, Pohang University of Science and Technology, Pohang, 790-784, Korea}	

\author	{H. Takatsu}
\affiliation{Department of Physics, Tokyo Metropolitan University, Tokyo 192-0397, Japan}
\affiliation{Department of Physics, Graduate School of Science, Kyoto University, Kyoto 606-8502, Japan}

\author{M. Hashimoto}
\affiliation{Stanford Synchrotron Radiation Lightsource, SLAC National Accelerator Laboratory, 2575 Sand Hill Road, Menlo Park, California 94025, USA}

\author{S.-K. Mo}
\author{Z. Hussain}
\affiliation{Advanced Light Source, Lawrence Berkeley National Laboratory, Berkeley, California 94720, USA}

\author{T. Oguchi}
\affiliation{Institute of Scientific and Industrial Research, Osaka University, 8-1 Mihogaoka, Ibaraki, Osaka 567-0047, Japan }

\author{T. Shishidou}
\affiliation{Department of Quantum Matter, ADSM, Hiroshima University, Higashi-Hiroshima 739-8530, Japan}

\author {Y. Maeno}
\affiliation{Department of Physics, Graduate School of Science, Kyoto University, Kyoto 606-8502, Japan}

\author     { B. I. Min }
\affiliation{ Department of Physics, Pohang University of Science and Technology, Pohang, 790-784, Korea}	
	
\author{Z.-X. Shen}
\email{zxshen@stanford.edu}
\affiliation{Stanford Institute for Materials and Energy Sciences, SLAC National Accelerator Laboratory, 2575 Sand Hill Road, Menlo Park, CA 94025, USA}
\affiliation{Geballe Laboratory for Advanced Materials, Departments of Physics and Applied Physics, Stanford University, Stanford, CA 94305, USA}

% ***** DATE
\date{\today}

% ***** ABSTRACT
\begin{abstract}
PdCrO$_2$ is material which has attracted interest due to the coexistence of metallic conductivity associated with  itinerant Pd $4d$ electrons and antiferromagnetic order arising from localized Cr spins.  A central issue is determining to what extent the magnetic order couples to the conduction electrons.  Here we perform angle-resolved photoemission spectroscopy (ARPES) to experimentally characterize the electronic structure.  We find that the Fermi surface has contributions from both bulk and surface states, which can be experimentally distinguished and theoretically verified by slab band structure calculations.  The bulk Fermi surface shows no signature of electronic reconstruction in the antiferromagnetic state.    This observation suggests that there is negligible interaction between the localized Cr spin structure and the itinerant Pd electrons measured by ARPES.
\end{abstract}

% ***** CONTENT
\maketitle
\section{Introduction}
Many strongly correlated electronic phases of matter exist in close proximity to antiferromagnetic (AFM) order, such as the high-temperature superconductivity of cuprates \cite{Lee2006} and iron pnictides,  \cite{Zhao2008} as well as the heavy fermion behavior of $f$-electron systems.\cite{Si2010}  In heavy fermion materials, for example, mobile conduction electrons form a bound resonance with localized magnetic moments, causing the electron effective mass to increase by a factor of $\sim$1000 as compared to conventional metals.  A complete understanding of these materials has proven elusive due to the competition or coexistence of several phases, such as paramagnetism, antiferromagnetism, and unconventional superconductivity.\cite{Lee2006,Zhao2008,Si2010}  It is therefore of interest to study the interaction of mobile conduction electrons and localized magnetic moments in a less complicated system where the physics may be more easily understood.

PdCrO$_2$ is material which has attracted attention due to the rich physics of its geometrically frustrated spin system.\cite{Takatsu2009a,Takatsu2009,Takatsu2010a,Takatsu2010}  The material crystallizes in a delafossite structure with $R\bar{3}m$ symmetry, and can be described as alternating stacked layers of Pd and Cr triangular lattices.  The localized spins of the Cr$^{3+}$ ions ($S=3/2$) exhibit AFM order at $T_\textrm{N} = 37.5$~K by forming a 120$^{\circ}$ spin structure with $\sqrt{3}\times\sqrt{3}$ periodicity.\cite{Mekata1995}  The material also exhibits metallic conductivity due to the Pd $4d^9$ electrons.\cite{Takatsu2009}  The physical properties of PdCrO$_2$ are analogous to those of the isostructural compound PdCoO$_2$, which also has metallic conductivity due to its Pd $4d^9$ electrons.\cite{Tanaka1998,Hicks2012}  The replacement of Co with Cr, however, introduces localized spins, and thus provides a system in which to study the coexistence of magnetic order and metallic conduction.\cite{Takatsu2009a,Ong2011}  Moreover, frustration effects arising from the geometric arrangement of the spins have unique signatures in transport measurements, lead to unusual critical exponent behavior,\cite{Takatsu2009} and also play a role in the recent observation of the unconventional anomalous Hall effect. \cite{Takatsu2010}
 
\begin{figure}[t]
\resizebox{\columnwidth}{!}{\includegraphics{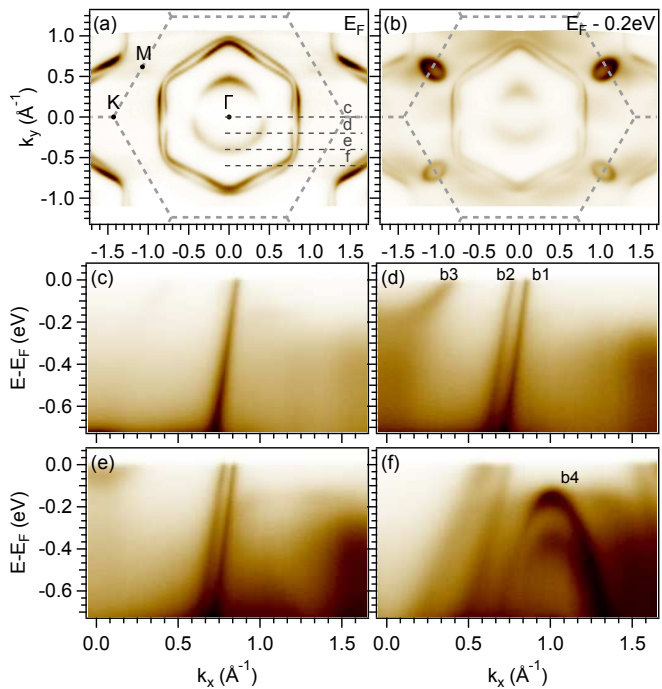}}
\caption{(Color online) (a) Constant-energy map of ARPES intensity at $E_{\textrm{F}}$, showing three FS pockets.  The data is symmetrized about $k_x = 0$.  The BZ boundaries and high-symmetry points are indicated.  The contours labeled c-f represent the cuts shown in the subsequent panels. (b) Constant-energy map at a binding energy of 0.2~eV.  Hole bands centered at the M-point are observed.  (c-f) Energy-momentum cuts at the momenta specified in panel (a).  The four bands discussed in the text are labeled b$_1$ through b$_4$.
\label{fig1}}
\end{figure}

\begin{figure}
\resizebox{\columnwidth}{!}{\includegraphics{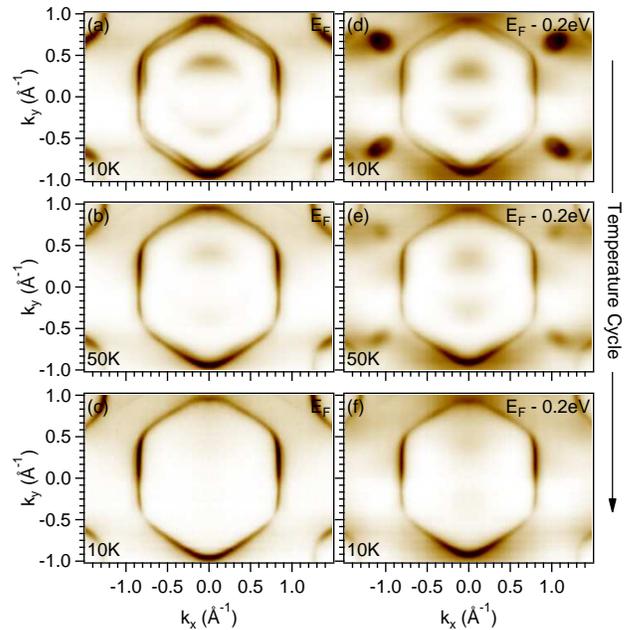}}
\caption{(Color online) Constant-energy maps during the temperature cycling process.  (a), (b), (c) are taken at $E_{\textrm{F}}$ at 10~K, 50~K, and 10~K, sequentially, on the same sample.  (d),(e),(f) The corresponding maps at $E_{\textrm{F}}-0.2$~eV. Note the disappearance of bands b$_2$, b$_3$, and b$_4$ during the cycling process.
\label{fig2}}
\end{figure}

Here we report an angle-resolved photoemission spectroscopy (ARPES) and band structure calculation study of the electronic structure of PdCrO$_2$ both above and below $T_\textrm{N}$.  The bulk Fermi surface (FS) is found to consist of a single hexagonal electron pocket.  In addition, surface states corresponding to a Pd-terminated crystal are identified.  We detect no signatures of electronic reconstruction in the AFM state.  This experimental observation suggests that the AFM order plays a negligible role in shaping the electronic structure of the itinerant Pd electrons measured by ARPES.

\section{Methods}
High-quality single crystals of PdCrO$_2$ were grown by a flux method and characterized with powder x-ray diffraction and energy dispersive x-ray analysis.\cite{Takatsu2010b}  The ARPES measurements were performed at beamline 10.0.1 of the Advanced Light Source using linearly polarized photons of energy 55~eV.  A Scienta R4000 hemispherical analyzer was used, with a total energy resolution of $\sim$20~meV and an angle resolution of $\sim$0.25$^{\circ}$, corresponding to a momentum resolution of 0.005~\AA$^{-1}$.  The samples were cleaved $in-situ$ at a base pressure $<5\times10^{-11}$~torr.

For bulk band structure calculations, we have used the full potential linearized augmented plane wave (FLAPW) method with local orbital (FLAPW+LO, APW+lo+LO), implemented in 
Wien2k\cite{Blaha2001} in the generalized gradient approximation (GGA).  Since our ARPES data includes contributions from surface states, we have additionally performed slab band structure calculations.  These calculations simulate the broken translational symmetry at the sample surface by employing a superstructure of stacked layers separated by a vacuum region.  For our calculations, we considered hexagonal unit cells separated by about 20~\AA~ of vacuum along the $z$-axis.  Each hexagonal unit cell contains 6 Cr layers, which was found to be sufficient to screen the surface effect and reasonably reproduce the bulk band structure. The vacuum dimension is large enough so that there is effectively	no hybridization between surface atoms through the vacuum. 

\section{Results}

In Figs. 1(a) and (b), we show constant-energy maps of the ARPES intensity at the Fermi energy $E_{\textrm{F}}$ and at $E_{\textrm{F}}-0.2$~eV, respectively, at a measurement temperature of 10~K.  There are three electron-like FS pockets, all centered at the $\Gamma$-point, with dispersions as shown in the energy-momentum cuts of Figs.1(c)-(f).  Bands b$_1$ and b$_2$ have hexagonal FSs and disperse linearly, while b$_3$ disperses parabolically and has an essentially circular FS (Fig. 1(a)).  The intensity of b$_2$ and b$_3$ are suppressed around $k_y=0$ due to photoemission matrix elements.\cite{Damascelli2003}  In addition to these three bands which cross $E_{\textrm{F}}$, there exist hole bands b$_4$ centered at the M-point at $\sim0.2$~eV binding energy.  The ARPES intensity is periodic with respect to the Brillouin zone (BZ) boundaries (indicated by grey dashed lines in Figs. 1(a) and (b)), which is consistent with the lattice's hexagonal symmetry.  From the periodicity we estimate a lattice constant of 2.97$\pm$0.06~\AA, which is consistent with the known value of the bulk lattice constant of 2.93~\AA.\cite{Takatsu2009} 

We now show that these bands can be experimentally determined to consist of bulk and surface states.  This insight is gained by cycling the sample temperature.  Specifically, we raised the sample temperature from 10~K to 50~K, and subsequently lowered the temperature back to 10~K.   During the warming process the chamber pressure temporarily increased up to $\sim1\times10^{-10}$~torr due to outgassing from the sample manipulator.  Constant-energy maps at the elevated and lowered temperatures are displayed in Fig. 2.  The most striking observation is the disappearance of intensity from bands b$_2$, b$_3$, and b$_4$ during the warming process, which does not recover after cooling the sample back to low temperature.  This irreversibility suggests that the change is due to sample surface contamination from the degassing process.  The disappearance of bands b$_2$, b$_3$, and b$_4$ indicates that they originate from surface states, while the surviving pocket b$_1$ is a bulk state.  (We note that the same thermal-cycling procedure has been successfully applied to distinguish surface and bulk states in PdCoO$_2$\cite{Noh2009} and related compounds.\cite{Yang2005})  The b$_1$ pocket has an area of 45$\pm$6\% of the BZ, and can therefore be attributed to one Pd 4$d$ conduction electron per unit cell.   

\begin{figure}
\resizebox{\columnwidth}{!}{\includegraphics{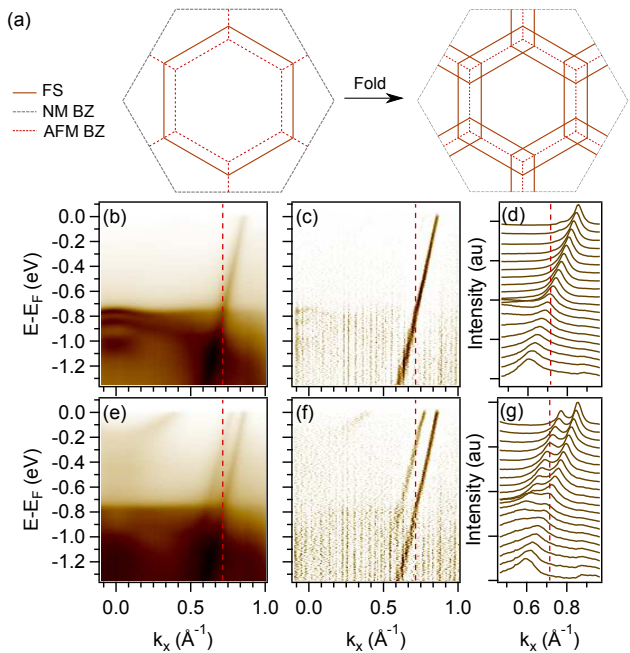}}
\caption{(Color online) (a) Cartoon of the FS reconstruction expected in the AFM state.  (b) ARPES cut at $k_y=0$ at $T=10$~K.  The red dashed line represents the AFM BZ boundary.  (c) Second derivative of the cut along $k_x$, helping to remove background and emphasize weak spectral features.  No folded bands are observed.  (d) The corresponding momentum distribution curves.  (e-g) The same information for the cut along $k_y=-0.2$~\AA.
\label{fig3}}
\end{figure}

\begin{figure}
\resizebox{\columnwidth}{!}{\includegraphics{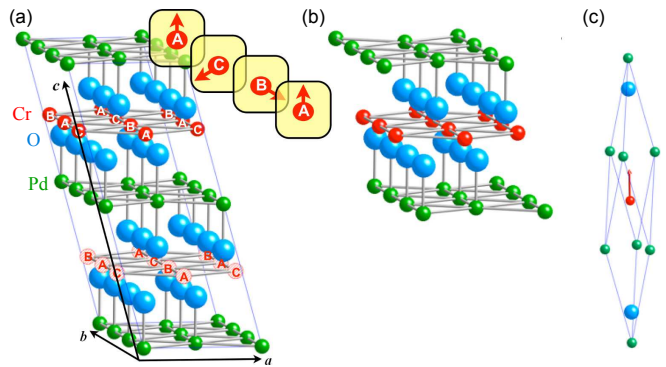}}
\caption{(Color online) The different spin configurations considered for the band structure calculations in this work:  (a) AFM order.  The Cr spins exhibit 120$^{\circ}$ structure with $\sqrt{3}\times\sqrt{3}$ periodicity.  The $c$-axis coupling between Cr layers is antiparallel.  (b) Same, but with parallel $c$-axis coupling.  (c) FM order.  The FM unit cell is identical to the NM unit cell.
\label{fig4}}
\end{figure}

We note that the geometry of the bulk FS b$_1$ is unchanged between 10~K and 50~K (Figs. 2(b) and (c)), which are temperatures below and above $T_\textrm{N}$ respectively.  This comparison suggests that the AFM order plays a negligible role in shaping the electronic structure.  This is surprising because the $\sqrt{3} \times \sqrt{3}$ AFM spin structure reduces the non-magnetic (NM) BZ to an AFM BZ with $\frac{1}{3}$ the area, and coupling to this reduced periodicity should lead to a corresponding reconstruction of the band structure,\cite{Voit2000} as illustrated in Fig. 3(a). Although no such folding is immediately observed, it is prudent to inspect the data carefully since the so-called shadow bands could have weaker spectral weight than the non-reconstructed bands. \cite{Voit2000,Brouet2004}

In Fig. 3(b) we show a cut through $k_y=0$, equivalent to Fig. 1(c) but plotted over a larger energy scale, at a measurement temperature 10~K $<$ $T_\textrm{N}$.  In the presence of AFM reconstruction the band b$_1$ would fold across the AFM BZ boundary, represented by the red dashed line.   In Fig. 3(c) we plot the 2$^\textrm{nd}$ derivative along $k_x$, which serves to remove slowly varying background features such as the intense valence states at 750~meV binding energy, and is therefore highly sensitive to weak spectral features.  Even after this procedure there is no indication of band folding across the AFM BZ boundary.  A close inspection of the momentum distribution curves (Fig. 3(d)) is consistent with the absence of band folding.  The same analysis is shown in Figs. 3(e)-(g) for $k_y=-0.2$~\AA, corresponding to the cut in Fig. 1(d).  This cut contains b$_2$ in addition to b$_1$, but again no reconstruction of either band is observed.  We therefore conclude that the ARPES data contains no signatures of reconstruction in the AFM state of PdCrO$_2$.

To better understand the electronic structure we have performed a number of band structure calculations.  We begin by computing the bulk band structure with AFM spin configuration, which reflects the magnetic ordering of the material below $T_\textrm{N}$.  Since the interlayer magnetic coupling is not known, we considered both antiparallel and parallel coupling between the Cr layers, as illustrated in Figs. 4(a) and (b).  We find that both configurations are degenerate, suggesting a weak Cr-Cr interlayer coupling.  The calculated FS for parallel interlayer coupling is shown in Fig. 5(a), and can be understood as originating from a single hexagonal pocket in the NM BZ which is then folded into the AFM BZ.  This understanding is consistent with the cartoon picture of Fig. 3(a).  However, when comparing to the experimental data, this result has two discrepancies: (1) As discussed above, the experimental data shows no indication of band folding due to the AFM order, and (2) The experimental data includes contributions from surface states, which are absent in these calculation results.

\begin{figure}
\resizebox{2.5in}{!}{\includegraphics{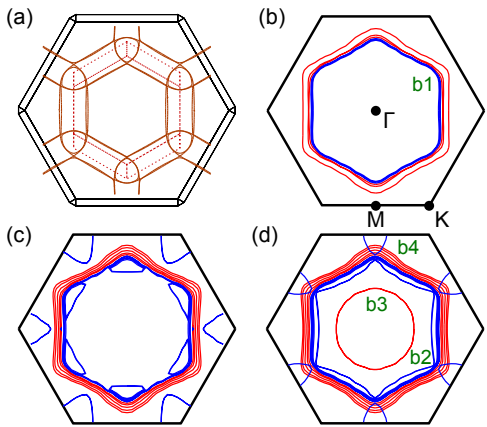}}
\caption{(Color online) FS of PdCrO$_2$ from band structure calculations.  (a) Bulk calculation with AFM spin configuration and parallel interlayer coupling.  The outer black hexagon represents the NM BZ boundary, while the red dashed lines mark the AFM BZ boundary.  The structure can be understood as a single hexagonal FS in the NM BZ which is folded into the AFM BZ. (b) Bulk calculation with FM spin configuration.  No folding is observed due to the equivalence of the FM and NM BZs.  The resulting FS agrees well with experiment, except for a small splitting due to the FM order (red/blue colors represent up/down spins).  (c) Slab calculation with O-terminated surface.  (d) Slab calculation with Pd-terminated surface.  The Pd-terminated slab calculation shows good agreement with experiment.
\label{fig5}}
\end{figure}

The first discrepancy is a fundamental consequence of the $\sqrt{3} \times \sqrt{3}$ periodicity of the AFM configuration used in the calculation.  To obtain a calculation result which resembles the experimental data, we must instead choose a configuration in which the periodicity is equivalent to that of the NM unit cell.  The simplest model with this periodicity, which also maintains the same local (polarized) spin structure as the AFM state, is a ferromagnetic (FM) spin configuration, illustrated in Fig. 4(c).  The calculated FS for this configuration is shown in Fig. 5(b).  As expected, the equivalence of the FM and NM unit cells implies that the electronic structure undergoes no reconstruction.  In fact, the resulting hexagonal pocket agrees well with b$_1$ observed in ARPES, except for a small spin-splitting arising from the FM order (red and blue colors in Fig. 5(b)). 

The second discrepancy, the absence of surface states, can be overcome by performing a slab band structure calculation.  This type of calculation simulates a sample surface by employing a superstructure of stacked layers separated by a vacuum region.  We continue to consider FM order to avoid reconstruction effects as discussed above.   Both Pd- and O- terminated surface conditions have been considered.  The calculation results are shown in Figs. 5(c) and (d).  In both cases, a number of FS pockets in addition to the bulk hexagon are observed, and can be identified as surface states.  In the O-terminated case, there are small FS pockets in the corners of the hexagon of b$_1$, as well as pockets centered at K, neither of which were observed in the experiment.  In the Pd-terminated case, a small inner-hexagon and circular electron pockets are observed, which can be identified with bands b$_2$ and b$_3$ observed in ARPES.  The calculation reveals an additional hole pocket centered at the M-point, which is not observed experimentally.  However, this pocket resembles the M-point hole bands observed at higher binding energy (see Fig. 1(b)), so we identify this state with b$_4$.  It is not clear why the binding energy of this band in the calculation disagrees so severely with experiment, but it may be associated with the fact that the FM spin configuration used in these calculations does not represent the true AFM order of the sample.  

\section{Discussion}

Using ARPES we have experimentally characterized the electronic structure of PdCrO$_2$.  The bulk FS is found to consist of a single hexagonal electron pocket which is attributed to one Pd 4$d$ conduction electron per unit cell.  This electronic structure bears a strong resemblance to that of PdCoO$_2$, which also consists of a single hexagonal electron pocket from its Pd 4$d$ electrons.\cite{Noh2009}  In addition, we observe electron pockets attributed to surface states.  From the slab band structure calculations we can identify that these states originate from a Pd-terminated surface.  This contrasts the situation of PdCoO$_2$, in which ARPES reveals features associated with a O-terminated surface, but not a Pd- terminated surface.\cite{Noh2009,Kim2009}  It is not understood why these two similar crystal structures should favor different surface terminations.

The deeper issue is the discrepancy between the non-reconstructed bands measured in ARPES and the reconstructed band structure expected for AFM order.\cite{Ong2011}  There are a number of examples of ARPES resolving band structure reconstruction in systems modulated by periodic potentials.  In the CeTe$_3$ system, for example, upon entering the charge density wave state the FS becomes gapped where it intersects the ordering vector, with a concomitant formation of folded shadow bands. \cite{Brouet2004}  Similar observations of gapping and folded shadow bands have been made in the spin density wave state of (Ba,Sr)Fe$_2$As$_2$. \cite{Yi2009}  Here we believe it is important to note a significant distinction between the density waves in these examples and the AFM state of PdCrO$_2$.  In these examples, the electrons associated with the reconstructed bands directly participate in the ordering.  For example, the gapped electrons in CeTe$_3$ are precisely those whose densities are modulated by the charge density wave.  In fact, the band reconstruction is thought to be a driving force for the density waves by providing a mechanism by which to lower the electronic kinetic energy.\cite{Schmitt2011}  Similarly, in the (Ba,Sr)Fe$_2$As$_2$ system it is believed that the itinerant electrons directly participate in the magnetic order.\cite{Hu2008} The situation is quite different in PdCrO$_2$, where the Pd conduction electrons themselves are not responsible for the order, but merely coexist with the AFM order coming from the Cr moments.\cite{Takatsu2009a} It is therefore not clear to what extent the conduction electrons should experience the periodic potential of the magnetic moments.  Since the spectral weight of the folded shadow bands is expected to be proportional to this coupling potential, \cite{Brouet2004, Voit2000} the absence of shadow bands may be taken as experimental evidence for weak coupling between the Pd 4$d$ conduction electrons and the localized Cr spins.  A similar example is found in EuFe$_2$As$_2$, where Eu contributes localized, ordered magnetic moments, and Fe$_2$As$_2$ contributes itinerant conduction electrons.  In this material, it is found that the low-energy electronic structure is nearly decoupled from the localized magnetic order.  \cite{Jeevan2008} 

Finally, we wish to acknowledge recent  observations of de Haas-van Alphen (dHvA) oscillations in PdCrO$_2$, which suggest the presence of several small FS pockets consistent with FS reconstruction in the AFM state. \cite{Ok2013}  This seeming contradiction between ARPES and dHvA results is strongly reminiscent of the case for underdoped cuprates, where ARPES reports a large FS or open arcs, \cite{Damascelli2003} but dHvA oscillations report small, closed FS pockets. \cite{Doiron-Leyraud2007a}  This discrepancy is not understood and is the subject of intense debate.  In our case, one possible interpretation is based on the surface-sensitivity of ARPES as compared to the bulk-sensitivity of dHvA experiments.  From the universal curve for inelastic mean free path of electrons in solids, we estimate that the 55~eV photons probe a sample depth of $\sim$0.5~nm. \cite{Seah1979}  This is substantially less than the $c$-axis lattice parameter of 1.8~nm and implies that our technique is highly surface sensitive.\cite{Takatsu2009}   While the bulk of the sample is known to have AFM order from transport and neutron scattering experiments, \cite{Takatsu2009} it is not known whether this order necessarily persists up to the surface.  The absence of shadow bands could therefore be attributed to short-range or non-existent AFM order near the surface of the sample.  On the other hand, another possible explanation is that the high magnetic fields used in dHvA experiments perturb the ground state of the material and lead to reconstructions which do not exist in the zero-field state probed by ARPES.  Similar explanations have been proposed for the cuprate experiments. \cite{Chen2008}  Further studies of dHvA oscillations, neutron scattering, and low photon energy ARPES (to achieve deeper bulk sensitivity) may be instrumental in resolving this issue.

\section{Conclusions}
 
We performed angle-resolved photoemission spectroscopy experiments on single-crystal samples of the compound PdCrO$_2$, which consists of alternating stacked layers of Pd and Cr triangular lattices.  We observed a large hexagonal bulk Fermi surface  centered at the $\Gamma$-point, attributed to the Pd $4d$ conduction electrons, as well as three surface-derived bands associated with a Pd-terminated crystal surface.  There was no electronic reconstruction observed with respect to the magnetic ordering of the Cr spins at $T_\textrm{N}$ = 37.5~K.  This result suggests that the Pd conduction electrons are weakly coupled to the magnetic order of the Cr spins.  Further investigations such as de Haas-van Alphen oscillation measurements  and neutron scattering experiments are important for further clarifying the interplay between the Pd conduction electrons and the Cr localized spins.  

\begin{acknowledgments}
We acknowledge P.~S. Kirchmann, R.~G. Moore, I.~M. Vishik, and S. Yang for fruitful discussions and experimental assistance.  The work at SLAC, Stanford, and LBNL is supported by the Department of Energy, Office of Basic Energy Sciences, Division of Materials Science.  JAS acknowledges support by the Stanford Graduate Fellowship.  KK and BIM acknowledge support by the NRF (Grants No. 2009-0079947 and No. 2011-0025237). HT and YM acknowledge support by MEXT KAKENHI (Grant No. 21340100), HT also by MEXT KAKENHI (Grant No. 24740240).    
\end{acknowledgments}

\bibliography{BibTex}
\end{document}